\magnification\magstep1
\vsize=7.5in
\hsize=5in

\def\sumint{\mathop{\rlap{$\sum\null$}\mkern-2mu\int}\nolimits}
\def\pp{\ \ .}
\baselineskip 12pt
\tolerance=1000
\widowpenalty=10000

\input BoxedEPS
\SetRokickiEPSFSpecial  
\HideDisplacementBoxes

\font\sc=cmcsc10 at 10pt
\nopagenumbers
\def\pagenumbers{\footline={\hss\tenrm\folio\hss}}
\pageno=0
\parindent=1.5em
\topglue1in
\centerline{\bf EFFECTS OF DIRAC'S NEGATIVE ENERGY}
\centerline{\bf SEA ON QUANTUM
NUMBERS}

\vskip 24pt
\centerline{R. Jackiw}
\vskip 24pt
\centerline{\it Center for Theoretical Physics}
\centerline{\it Laboratory for Nuclear Science}
\centerline{\it and Department of Physics}
\centerline{\it Massachusetts Institute of Technology}
\centerline{\it Cambridge, Massachusetts\ \ 02139\ \ U.S.A.}
\vfill\vfill\vfill
\centerline{Dirac Prize Lecture}
\centerline{Trieste, Italy -- March 1999}
\vfill
\eject

 \pagenumbers
\null\vskip.5in
\noindent
{\sc The Dirac Prize Committee} cited my work on fractional charges and on
chiral anomalies; therefore, I shall discuss these two topics. As with all true and
deep physical effects, there are many ways of arriving at the results. It is
particularly appropriate here today that one route towards understanding both
fractional charges and chiral anomalies delves into Dirac's negative energy sea.
This is especially provocative, because usually we think of Dirac's negative energy
sea as an unphysical construct, invented to render quantum field theory physically
 acceptable by hiding -- that is, by renormalizing -- negative energy
solutions. But I am suggesting that in fact physical consequences can be drawn
from Dirac's construction. 

First let me set the stage for the discussion.

Quantum  physics has taught us that a physical observable need not be a 
quantity with arbitrary magnitude.  Because it is the eigenvalue of a linear 
Hermitian operator, it will in general be quantized.  This is not 
the case in classical physics where most observables, like angular momentum 
and energy, are continuously varying and can take on any value.  On the other
hand, even in classical physics, there are concepts that  are intrinsically integral --
for example particle number of conserved  particles -- and one expects that the
integrality will be preserved in the  quantum theory, that is, eigenvalues of the
relevant operator -- the  number operator in our example -- are expected to be
integers.  Quantization of  eigenvalues is most easily attained when the operator is
a generator of a  compact, non-Abelian group, like angular momentum generating
SO(3)  rotations.  However, the number operator frequently generates only
Abelian  transformations with no group-theoretic quantization.

Closer examination of the number operator in a theory with second 
quantized fermions raises doubts that it will in fact 
possess only integer eigenvalues. 
 The problem derives from Dirac's negative energy sea, which must be filled 
to define the vacuum.  This involves an infinite number of ``particles''.  
Since the number of any further particles 
must be measured relative to this infinity, there 
may very well emerge a non-integral answer.  Nevertheless, it had been 
believed that 
various renormalization procedures, like normal ordering, can unambiguously 
ensure integrality of the eigenvalue.  Therefore, it was a suprise when Rebbi and I 
established about twenty years ago that fermions moving in the field of a 
topologically nontrivial soliton (kink in one spatial dimension, 
vortex in two, monopole in three) possess non-integer eigenvalues for their 
number operator.$^1$  It is perhaps even more surprising that this peculiar effect 
has a physical realization in properties of actual condensed matter 
systems -- polyacetylene being the standard example.$^2$  Here, I shall 
describe this to you, first in a general, formal way, and then in a 
physically intuitive language appropriate to polyacetylene.

We wish to second quantize fermions moving in a static 
background that generically 
is described by 
$\varphi$.  Fermion dynamics is governed by a Dirac Hamiltonian $H(\varphi)$.  
Two different 
backgrounds are envisioned:  one is appropriate for the vacuum sector of the 
theory $\varphi_v$, the other for the soliton $\varphi_s$.  For example, $\varphi$ 
may 
be a condensate field that takes a homogenous value in the vacuum sector and 
a topologically nontrivial profile in the soliton sector.  

\headline={\hfill{\it Effects of Dirac's Negative Energy Sea on Quantum
Numbers}\hfill}

Second quantization
proceeds by computing the energy eigenvalues and eigenfunctions of $H(\varphi)$, 
which possesses both positive and negative energy eigenstates, and 
``filling'' the negative energy sea.  The number density of the soliton ground 
state is given by $$\rho({\bf x}) =
\sumint^{0^-}_{-\infty} \!\!\! dE\, 
\left(|\Psi_E({\bf 
x})|^2 - |\psi_E({\bf x})|^2\right) \eqno(1)$$
where the integration, which also includes summutation over discrete levels, 
extends over all negative energy states, since they are filled in the vacuum.  
Here $\Psi_E$ is the energy eigenfunction in the presence of the soliton and 
$\psi_E$ is the eigenfunction in the vacuum sector:
$$
H(\varphi_s)\Psi_E = E\Psi_E,\quad  H(\varphi_v) \psi_E = E\psi_E
\pp
\eqno(2)
$$  
In (1) the contribution from the vacuum sector is subtracted; the soliton charge
density is renormalized, so that it is measured relative to the
vacuum.  The fermion number of the soliton ground state is the spatial integral of
$\rho$ 
$$
N_F = \int d{\bf x}\, 
\rho({\bf x}) \pp
\eqno(3)
$$

A very beautiful aspect of the theory is that one can evaluate (1) 
and (3) by general methods, which bypass solving 
the eigenvalue problem (2) explicitly.  Rather, one uses spectral sum 
rules whose form is dictated by general features of the Hamiltonian, 
in particular by the topological properties of the background $\varphi$ and of 
the space $\{{\bf x}\}$.  While these methods are powerful, they are also 
technical, requiring much mathematical knowledge, so I shall not present them 
here.  However, when the Hamiltonian posseses one further property, which I shall
now describe, the sum  rules become trivial, and the result for $N_F$ is immediate.

We assume further that $H(\varphi)$ possesses a conjugation symmetry, taking  
positive energy states into negative energy states and {\it vice versa}; that is, we
assume  there exists an operator ${\cal C}$ that anticommutes with
$H(\varphi)$: 
${\cal C}^{-1}H{\cal C} = -H$. 
 One consequence of this is that the number density at $E$ is an even 
function of $E$: $|\Psi_E|^2 = | \Psi_{-E}|^2,\quad | \psi_E|^2 = 
| \psi_{-E}|^2$. 
A less obvious consequence is that in the soliton sector, there are always  
normalizable, discrete zero-energy modes
$$
H(\varphi_s) u_0 = 0,\quad 
\int d{\bf x}\, |u_0({\bf x})|^2 =1 \pp
\eqno(4)
$$
This fact may be seen by explicit solution of the eigenvalue problem, but it 
also follows from a general mathematical argument, called {\it index theory}, 
which ensures that the Dirac operator has normalizable zero-energy modes in the 
presence of a topologically nontrivial background.

We are now in a position to evaluate (1) and (3).  First, we use completeness 
of the eigenfunctions in the soliton and vacuum sectors  
$$
\sumint^{0^-}_{-\infty}\!\!\!\!\!\!  dE\, |\Psi_E({\bf x})|^2 
+ \sumint^\infty_{0^+}\!\!\!  dE\, | \Psi_E({\bf x})|^2 
+ |u_0({\bf x})|^2 - \sumint^\infty_{-\infty}\!\!\!\!\!\!   dE\,  | \psi_E({\bf x})|^2 = 
0 \pp
\eqno(5)
$$
The zero-energy mode in the soliton sector has been explicitly separated; we 
assume there is just one.  (In the vacuum sector there are none.)  Then, 
use of the conjugation symmetry allows equating the positive energy 
integrals with the negative energy ones, and 
converts (5) into an evaluation of (1) 
$$
\sumint^{0^-}_{-\infty}\!\!\!   
dE\,  \left(| 
\Psi_E({\bf x})|^2 - | \psi_E({\bf x})|^2\right) = -{\textstyle{1\over 2}} |u_0({\bf 
x})|^2 \pp
\eqno(6)
$$
The spatial integration that determines $N_F$ is trivial 
since the zero mode is normalized:
$$
N_F = -{\textstyle{1\over 2}} \pp
\eqno(7)
$$
The conclusion is that the 
soliton vacuum, defined with the zero mode empty, carries fermion number 
$-{1\over 2}$; of course, when the zero mode is filled, this fermion number is 
$+{1\over 2}$!  The fermion number 
assignment of $\pm{1\over 2}$ for two states degenerate in 
energy is the only possible one consistent with a conjugation-odd fermion 
number operator.  

Two comments should be made in connection with this very elementary 
derivation of our surprising result.
\medskip
\item{(i)}  The above evaluation concerns the {\it expected} value of the 
second-quantized, field theoretic number operator, $\hat N_F$.  However, one 
can show, by expanding the second quantized field in terms of creation 
and annihilation operators in the presence of the soliton, that in fact 
the {\it eigenvalues} are $\pm{1\over 2}$.
\smallskip
\item{(ii)}  We have viewed the soliton as an external field.  In a complete 
description, one must take the soliton's 
quantum dynamics into account.  Necessarily there will occur 
spontaneous symmetry 
breaking in the vacuum sector.  Calculations in the full theory can be 
carried out by Monte-Carlo methods, or by analytic techniques of the 
Born-Oppenheimer variety.
\medskip

The three ingredients necessary for fermion number fractionization --  
spontaneous symmetry breaking, solitons and fermions -- come together in a 
description of a physical system, polyacetylene.  This is a 
one-dimensional array of carbon atoms that can form one of two degenerate 
ground states.  The degeneracy arises from a spontaneous breaking of the 
right-left symmetry (Peierls instability) and manifests itself in an 
alteration of the bonding pattern, as illustrated in Fig.~1.
\midinsert $$
\BoxedEPSF{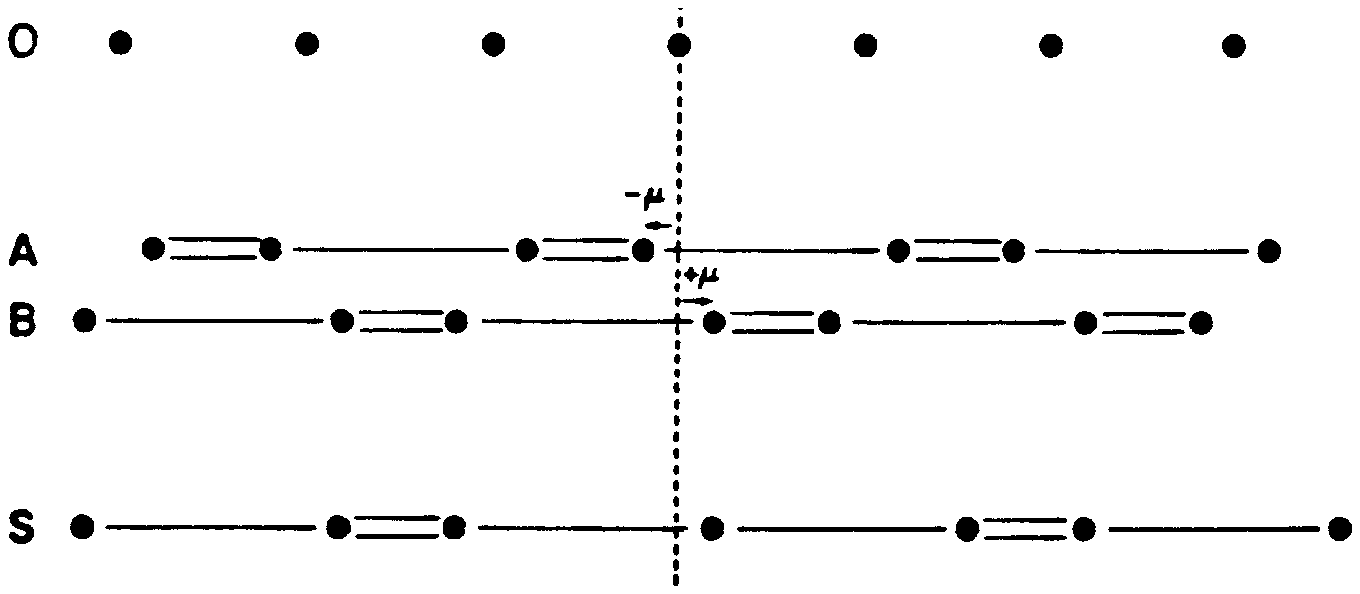 scaled 750}
$$ {\narrower\noindent Fig.~1: Polyacetylene consists of a
linear  chain of carbon atoms (dots).  The equally spaced configuration (O)
possesses  a left-right symmetry, which however is energetically unstable.  Rather
in the  ground states the carbon atoms shift a distance $\mu$ to the left or right, 
breaking the symmetry 
and producing two degenerate ground states (A,B).  (The drawing 
is not to scale; the shift is only 
a few percent of the total bond length.)  A 
soliton (S) is a defect in the alteration patters; it provides a domain wall 
between configurations (A) and (B).\par}\endinsert

A microscopic Hamiltonian for the system has been proposed by Su, Schrieffer 
and Heeger (SSH).$^2$  In the continuum and infinite volume limit, electron 
transport is governed by a Dirac-type Hamiltonian in one spatial dimension:
$$
\eqalign{
H(\varphi) &= \sigma_3 \hat p+ \sigma_1\varphi(x),
\quad \hat p={1\over i} \ 
{d\over dx} \cr \sigma_3 &= \Bigl(\matrix{1 & 0 \cr 0 & -1\cr}\Bigr),\quad 
\sigma_1 = \Bigl(\matrix{0 & 1\cr 1 & 0\cr}\Bigr) \pp
\cr} 
\eqno(8)
$$
Here, $\varphi(x)$ is the phonon field; it measures the displacement of the 
carbon atom from its equalibrium position.  The matrix structure of the above 
Hamiltonian does not arise from spin.  In the SSH description, 
electron-electron 
interactions are ignored and spin is a passive label; the Hamiltonian in (8) 
describes separately spin up and spin down electrons.  Rather, the Dirac-like
matrix form for $H$ arises through a linearized approximation and the 
two-component wavefunctions that are eigenmodes of $H$ refer to the 
right-moving and left-moving electrons with momentum
$\pm|p|$.  The filled  negative energy states are the valence electrons,
while the conduction  electrons populate the positive energy states.

In the SSH model, $\varphi(x)$ is 
determined self-consistently by the phonon's dynamics, and in the lowest 
(vacuum) states $\varphi(x)$ takes the uniform values $\pm\mu$, as illustrated 
in Fig.~1.  The corresponding spectrum of (8) exhibits a gap.  

In addition to the two ground states, where the phonon field takes a constant 
value, there exist stable excited states where $\varphi(x)$ assumes a kink 
shape, which interpolates as $x$ passes from $-\infty$ to $+\infty$, 
between the 
vacuum configurations $-\mu$ and $+\mu$.  This is the soliton, and it 
describes a defect in the alteration pattern, as is also exhibited in 
Fig.~1.

The Hamiltonian in (8) admits a conjugation symmetry:  ${\cal C}=\sigma_2 = 
\bigl({0 \atop i}{-i \atop 0}\bigr)$ anticommutes with $H$ -- this is 
ordinary charge conjugation invariance in the absence of Coulomb interactions. 
 The zero eigenvalue problem is easily solved with a kink 
background: there is one normalizable zero-energy solution.  Thus, our general 
analysis predicts that fermion number, here coinciding with charge, 
fractionizes to $\pm {1\over 2}$ in the one-soliton state.  

This result may also be 
seen pictorially.  When two solitons are inserted into the ground state $(B)$, 
the 
bonding pattern is depicted in Fig.~2.  Note that the number of bonds in the 
two-soliton state is one fewer than in the ground state.  
If the two solitons are 
now separated far apart, so that they act independently, the quantum numbers 
of the missing bond must be shared between the two states, and that is how the 
fraction ${1\over 2}$ arises.
\midinsert $$
\BoxedEPSF{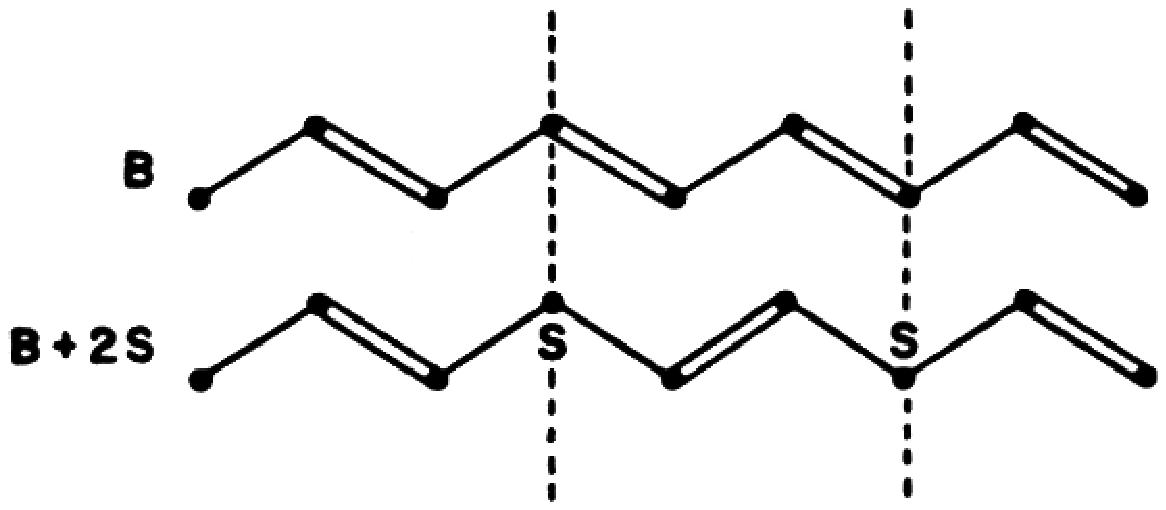 scaled 750}
$$  {\narrower\noindent Fig.~2: With two
solitons
(SS) inserted in  vacuum (B), the number of bonds $\approx$ electrons
between the sites of the defects decreases  from five to four.\par}\endinsert

Of course, for actual physical samples, where both the volume and the separation 
between defects are finite, the non-integer fermion number is only an 
expectation value for the operator $\hat N_F$, and there are corrections 
that vanish in the infinite limit.  The important point is that the variance 
$\langle\hat N^2_F\rangle - \langle \hat N_F\rangle^2$ also vanishes in the 
limit.  This is to be contrasted to the uninteresting situation of, say, 
an electron circulating 
about two nuclei.  When the 
nuclei are far apart, the expected value for electron number near each 
nucleus is ${1\over 2}$, 
plus small corrections.  However, the variance remains ${1\over 4}$ for 
infinite separation, which shows that this fraction never becomes an eigenvalue.

The concept of fractional quantum numbers has now extended beyond soliton 
systems -- for example the theory of the quantum Hall effect makes use of the 
idea, even though dynamical details are quite different from the above 
example.

While fractional quantum numbers were first seen in relativistic field 
theory, thus far they have not played any experimentally verified role in 
particle physics.  Nevertheless, it is curious that an effect which in 
principle is physical, and has been observed in  condensed matter
systems, should arise from distortions in the negative energy  sea, which for
particle physicists  is an unphysical construct, in contrast to the condensed matter
application,  where negative energy states correspond to physical quantities -- the
valence  electrons.

In particle 
physics there is another, physically realized circumstance  where the Dirac
negative energy sea modifies symmetry behavior of  fermions.  This is the chiral
anomaly phenomenon whereby the axial vector  currect $i\bar\psi \gamma^\mu
\gamma^5\psi$, which is conserved for free  massless Dirac fermions, ceases to be
conserved when the massless fermions are quantized and made to interact 
with a gauge field, even though the interaction appears to be chirally
invariant.$^{3}$  
Indeed, for unquantized Dirac fields the chiral invariance ensures conservation of
the unquantized axial vector current. This disappearance after quantization of
chiral symmetry is usually associated 
with infinities that plague relativistic quantum field theory:  
the infinities must be regulated and renormalized, but there is no chirally 
invariant regulator procedure.  However, a more directly physical discussion 
of the anomaly phenomenom may be given, which shows that in fact it is the 
filling of the negative energy sea that breaks the chiral symmetry.

Let me first state the essential puzzle of the chiral anomaly. 
In spite of profound differences between the classical and quantal description of
physical phenomena, it was generally believed that symmetry properties and
conserved constants of motion transcend the classical/quantal dichotomy: when a
classical model possesses symmetries and supports constants of motion, one
expects that quantization preserves the symmetries, so that conserved quantities
-- now propmoted to quantum operators -- remain time-independent, that is,
they commute with the Hamiltonian operator. But as observed thirty years ago by
Bell and me,$^{3}$ and also independently by Adler,$^{3}$ this need not be~so.

A simple instance of quantum mechanical violation of symmetry is encountered
by considering 
 a 
massless Dirac fermion moving in a background gauge field $A_\mu$.  
The dynamics is governed by a Lagrangian, which splits into separate right and 
left parts -- this separation is a manifestation of chiral symmetry:
$$
\eqalign{
{\cal L} &= \bar\psi(i\not{\mkern-3mu\partial} - e\not{\mkern-5mu A})\psi \cr 
&= \bar\psi_+(i \not{\mkern-3mu\partial} - e\not{\mkern-5mu A})\psi_+ +
\bar\psi_-(i\not{\mkern-3mu\partial} -  e\not{\mkern-5mu A})\psi_-\cr \psi &=
\psi_++\psi_-,\quad
\psi_{\pm} = {\textstyle{1\over 2}}(1
\pm  i\gamma_5)\psi\pp
\cr}
\eqno(10)
$$
In the first-quantized theory, where $\psi$ is a wavefunction and $\bar\psi 
\gamma^\mu\psi$ is a probability current, the separate right and left currents 
are conserved, and the separate probabilities $\int d{\bf x}\,  
\psi^\dagger_{\pm}\psi_{\pm}$ are time-independent.  In the second quantized 
theory, where $\psi$ becomes an operator, the anomaly phenomenon renders the 
separate right and left currents no longer conserved, and the right and left 
charges are not time-independent.  Nevertheless, the sum of right and left 
currents -- 
the vector current -- is conserved, while the divergence of the difference 
between the  
right and left currents -- the axial vector current -- is nonzero owing to the 
anomaly.  Our task then is to understand what causes the separate 
nonconservation of left and right currents even though there is no coupling 
between the two apparent in (10).  

Evidently, the problem derives 
from the second quantization procedure, hence 
we implement it.  We set $A^0$ to zero, find the eigenmodes of the Hamiltonian 
in the background field ${\bf A}$, and define the second quantized vacuum by 
filling the negative energy modes, leaving the positive energy modes 
empty.  The background ${\bf A}$ is chosen in a specific functional form so 
that the anomaly is nonvanishing.  This requires ${\bf 
A}$ be   time-dependent, but we 
chose a potential constant in 
time and space and model the time variation by an adiabatic change 
${\bf A}\to {\bf A}+\delta{\bf A}$.

The simplest model to study is two-dimensional and Abelian -- two-dimensional 
massles quantum electrodynamics.$^4$
The Dirac matrices are $2\times 2$ and $\psi$ is a two-component spinor.
$$\eqalign{
\gamma^0 &= \sigma^1,\quad \gamma^1 = i\sigma^2,\quad \gamma_5 = 
i\gamma^0\gamma^1=-i\sigma^3 \cr \psi_+ &= 
\Bigl(\matrix{1 & 0 \cr 0 & 0\cr}\Bigr)\psi,\quad \psi_- = \Bigl(\matrix{0 & 
0 \cr 0 & 1\cr}\Bigr)\psi \pp \cr}
\eqno(11)
$$
The axial current possess an anomalous divergence proportional to 
$\epsilon^{\mu\nu}F_{\mu\nu}\propto \partial_t A$ (since $A_0=0$ and
$\bf A$ has the single spatial component $A^1 \equiv A$),$^{5}$ which we wish to
understand. The eigenmodes to be  second quantized satisfy a one-dimensional
Dirac equation, 
$$
H\psi_E = 
\alpha(\hat p-eA)\psi_E = E\psi_E,\quad \alpha = -\sigma^3 
\eqno(12)
$$
where $A$ is constant.  They are given by 
$$\eqalign{\psi_+ &= 
\Bigl(\matrix{ e^{ipx} \cr 0\cr}\Bigr) \ {\rm with}\ E=-p+eA 
\cr\noalign{\vskip 0.2cm} \psi_- &= \Bigl(\matrix{0 \cr e^{ipx}\cr}\Bigr)\ 
{\rm with}\ E=p-eA\cr} \pp
\eqno(13)
$$

Second quantization is performed by filling the negative energy sea.  For 
$A=0$, the energy-momentum dispersion is depicted 
in Fig.~3, where the right-hand 
branch corresponds to fermions of one chirality, and the left-hand branch to 
those of the other chirality.  The negative energy states are filled, as 
indicated by the filled circles; the positive energy states are empty, as 
indicated by the empty circles.  As $A$ increases from 0 to $\delta A$, empty 
states in the right-hand branch acquire negative energy, while filled states 
of the left-hand branch become positive energy states; that is, there is a 
net production of right-handed antiparticles and left-handed particles; see 
Fig.~4.  So the separate right and left charges are not conserved, 
but their sum is.  Put in another way, the separation between positive and 
negative energy states of definite chirality cannot be achieved 
gauge-invariantly, since changing $A$ from 0 to a constant $\delta A$ is a 
gauge transformation,  yet particles are produced.

\midinsert 
\hfuzz=8pt
\vbox to 1.95in{
$$
{\BoxedEPSF{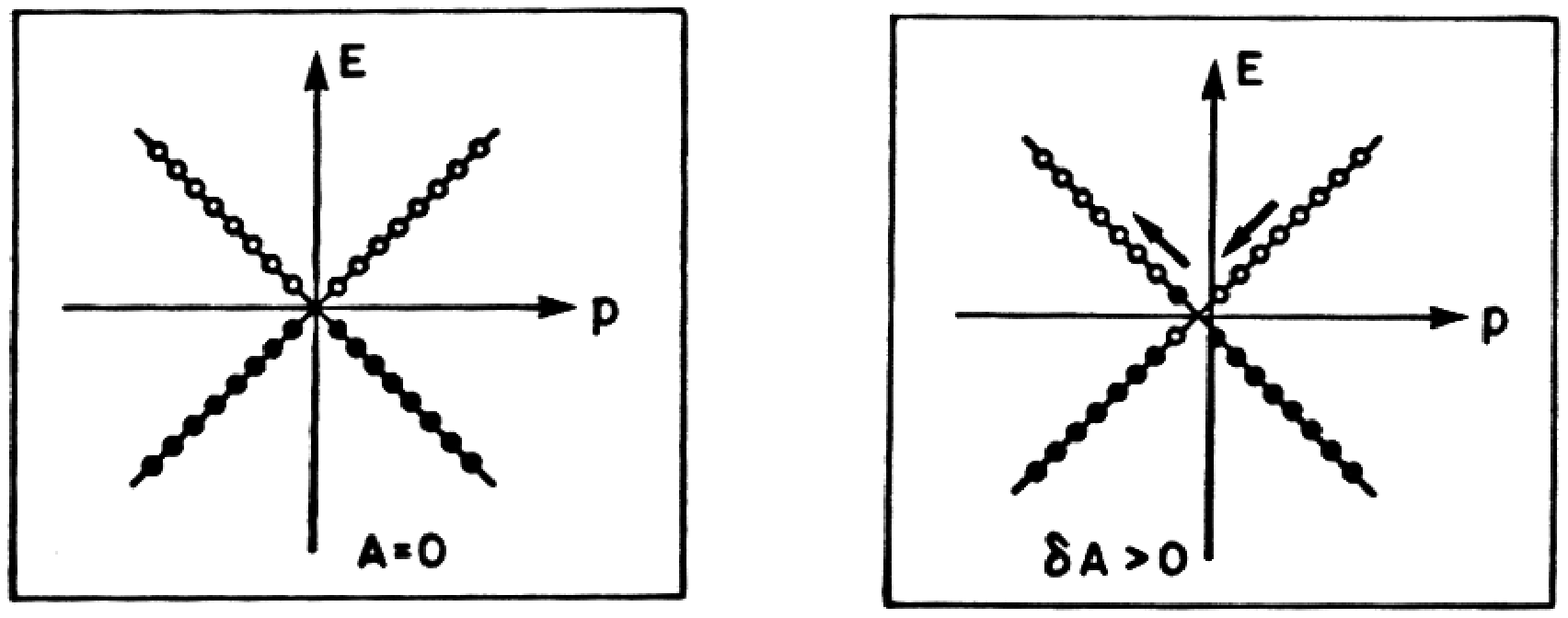 scaled 700}}
$$} \smallskip\hbox{\vtop{\hbox to
2.3in{\quad Fig.~3:  Energy-momentum disper-}\hbox to 2.3in{\quad sion at
$A=0$.   Empty circles are} \hbox to 
2.3in{\quad empty states; filled circles are filled}\hbox to 2.3in{\quad 
states.\hfil}}
\hskip .2in \vtop{\hbox to 2.3in{Fig.~4: Energy-momentum dispersion}
\hbox 
to 2.3in{at $A=\delta A>0$.  The energy shift\quad}
\hbox to 2.3in{produces negative energy empty states\ \ }
\hbox to 2.3in{and positive energy filled states.\hfil}}}
\smallskip\endinsert

We thus see that the negative energy sea is responsible for 
nonconservation of chirality even though the dynamics is chirally invariant.  
This effect was called {\it anomalous\/} because its discovery was a
surprise.$^{3}$  However, a better name might be {\it quantum mechanical
symmetry  breaking} -- a symmetry-breaking mechanism, which like Heisenberg's
{\it  spontaneous 
symmetry breaking\/} attributes physical asymmetry to the vacuum state and
not  to the dynamics.  Here, however, unlike in Heisenberg's case, it is not 
vacuum degeneracy but the very definition of the vacuum that is responsible.
Once again we must assign physical reality to Dirac's negative energy sea, because
it produces the chiral anomaly, whose effects are experimentally observed,
principally in the decay of the neutral pion to two photons, but there are other
physical consequences as well.$^6$

The two phenomena that I have described show that Nature seems to know and
make use of what at first appears to be a defect of a quantum field theory, and
which is usually ``renormalized away''.  Remarkably, the infinities in the
formalism give rise to finite and physical effects. One may quite appropriately call
this an example of ``The Unreasonable Effectiveness of Quantum Field Theory in
Physics''. 

In my presentation of fractional charge and of the chiral anomaly
I have used concepts that are primitively physical,
employing little analysis beyond drawing pictures and counting. But to the same
end a most sophisticated mathematical discussion can also be given, wherein zero
modes are controlled by various index theorems (Atiyah-Singer, Callias),
fractional charges are related to Atiyah, Patodi, Singer spectral flows, and chiral
anomalies are identified with Chern-Pontryagin densities. This confluence
between physics and mathematics, which was brought about by physicists'
research on the topics that I described, seeded an interaction between these two
disciplines, which still flourishes  and today fuels the current string/M-theory
program. 

But in sharp contrast with the above, contemporary mathematical
discussions within physics do not as yet have experimentally observed
correlatives. These days research  follows very closely Dirac's dictum:
\medskip{%
\narrower\noindent
The most powerful method of advance [in physics]\thinspace$\ldots$\thinspace is
to employ all the resources of pure mathematics in attempts to perfect and
generalize the mathematical formalism that forms the existing basis of theoretical
physics, and\thinspace$\ldots$\thinspace to try to interpret the new
mathematical features in terms of physical entities.\hfill --- {\it
P.A.M.~Dirac} 
\par}
\medskip\noindent
Early in my career I heard him say this at a seminar devoted to his research
history, and I was very inspired. However, my subsequent work with physics and
mathematics makes me feel that Dirac's suggestion is too radical, while a quote by
Yang more accurately describes my own experience of mathematical physics: 
 \medskip
\narrower\noindent{
Physics is not mathematics, just as mathematics is not physics. Somehow
nature chooses only a subset of the very beautiful and complex and intricate
mathematics that mathematicians develop, and that precise subset is what
the theoretical physicist is trying to look for.\hfill --- {\it C.N.~Yang} 
\par}
\vfill\eject

\centerline{\bf REFERENCES}
\frenchspacing 
\smallskip
\item{1.}  R. Jackiw and C. Rebbi, {\it Phys. Rev. D} {\bf 13}, 3398 (1976); for a
review, see A. Niemi and G. Semen\-off, {\it Phys. Reports} {\bf 135}, 99
(1986).
\medskip
\item{2.}  W.-P. Su, J.R. Schrieffer and A. Heeger, {\it Phys. Rev. Lett.} 
{\bf 42}, 1698 (1979) and {\it Phys. Rev. B} {\bf 22}, 2099 (1980).
\medskip
\item{3.}  J.S.~Bell and R. Jackiw, {\it Nuovo Cim.} {\bf 60}, 47 (1969); S.~Adler,  {\it
Phys. Rev.} {\bf 177}, 2426 (1969).
\medskip
\item{4.}  J. Schwinger, {\it Phys. Rev.} {\bf 128}, 2425 (1962).
\medskip
\item{5.}  K. Johnson, {\it Phys. Lett.} {\bf 5}, 253 (1963).
\par
\medskip
\item{6.}  For extensive discussion of the anomaly phenomena see S. Treiman, 
R. Jackiw, B. Zumino and E. Witten, {\it Current Algebra 
and Anomalies},  (Princeton University Press/World Scientific, Princeton, 
NJ/Singapore, 1985) or S.~Adler in {\it Lectures on Elementary Particles and
Quantum Field Theory}, S.~Deser, M.~Grisaru, and H.~Pendleton, eds.\ (MIT Press,
Cambridge, MA, 1970).

\bye